\documentclass[10pt,twocolumn,letterpaper]{article}

\usepackage{cvpr}              

\usepackage[accsupp]{axessibility}  
\usepackage{graphicx}
\usepackage{amsmath}
\usepackage{amssymb}
\usepackage{booktabs}
\usepackage{multirow}
\usepackage{paralist}
\usepackage{array}
\usepackage[normalem]{ulem}
\usepackage[accsupp]{axessibility}
\usepackage{soul}
\usepackage{url}

\usepackage{kotex}

\usepackage{subcaption}

\usepackage[pagebackref,breaklinks,colorlinks,bookmarks=false]{hyperref}

\usepackage[capitalize]{cleveref}
\crefname{section}{Sec.}{Secs.}
\Crefname{section}{Section}{Sections}
\Crefname{table}{Table}{Tables}
\crefname{table}{Tab.}{Tabs.}

\usepackage{algorithm}
\usepackage[noend]{algpseudocode}

\usepackage{chngcntr}

\def\cvprPaperID{8858} 
\def\confName{CVPR}
\def\confYear{2023}

\usepackage{dsfont}
\usepackage{stix}

\usepackage[usenames,dvipsnames]{xcolor}

\newcommand{\blockcomment}[1]{}

\newcommand{\ours}{OCELOT\xspace}
\newcommand{\supple}{supplementary material\xspace}
\newcommand{\score}[2]{{#1}\scriptsize{$\pm$#2}}

\makeatother

\begin{document}

\title{OCELOT: Overlapped Cell on Tissue Dataset for Histopathology}


\author{
Jeongun Ryu\thanks{: Equal contribution} \hspace{0.4mm}
Aaron Valero Puche\footnotemark[1] \hspace{0.4mm} 
JaeWoong Shin\footnotemark[1] \hspace{0.4mm}
Seonwook Park \hspace{0.4mm}
Biagio Brattoli \hspace{0.4mm}
Jinhee Lee 
\\
Wonkyung Jung \hspace{0.4mm}
Soo Ick Cho \hspace{0.4mm}
Kyunghyun Paeng \hspace{0.4mm}
Chan-Young Ock \hspace{0.4mm}
Donggeun Yoo \hspace{0.4mm}
Sérgio Pereira
\\[1mm]
Lunit Inc. \\[0.5mm]
\tt\small \{rjw0205, aaron.valero, jwoong.shin, spark, biagio, jinhee.lee,
\\
\tt\small wkjung, sooickcho, khpaeng, ock.chanyoung, dgyoo, sergio\}@lunit.io
\vspace{-3mm}
}
\maketitle

\vspace{-3mm}
\begin{abstract}
Cell detection is a fundamental task in computational pathology that can be used for extracting high-level medical information from whole-slide images.
For accurate cell detection, pathologists often zoom out to understand the tissue-level structures and zoom in to classify cells based on their morphology and the surrounding context.
However, there is a lack of efforts to reflect such behaviors by pathologists in the cell detection models, mainly due to the lack of datasets containing both cell and tissue annotations with overlapping regions.
To overcome this limitation, we propose and publicly release \ours, a dataset purposely dedicated to the study of cell-tissue relationships for cell detection in histopathology. \ours provides overlapping cell and tissue annotations on images acquired from multiple organs.
Within this setting, we also propose multi-task learning approaches that benefit from learning both cell and tissue tasks simultaneously. 
When compared against a model trained only for the cell detection task, our proposed approaches improve cell detection performance on 3 datasets: proposed \ours, public TIGER, and internal CARP datasets.
On the \ours test set in particular, we show up to $6.79$ improvement in F1-score.
We believe the contributions of this paper, including the release of the \ours dataset at \url{https://lunit-io.github.io/research/publications/ocelot} are a crucial starting point toward the important research direction of incorporating cell-tissue relationships in computation pathology.
\vspace{-5mm}
\end{abstract}


\section{Introduction}
\label{sec:intro}

\begin{figure*}[t]
    \centering
    \hfill
    \vskip -3mm
        \includegraphics[width=0.8\linewidth]{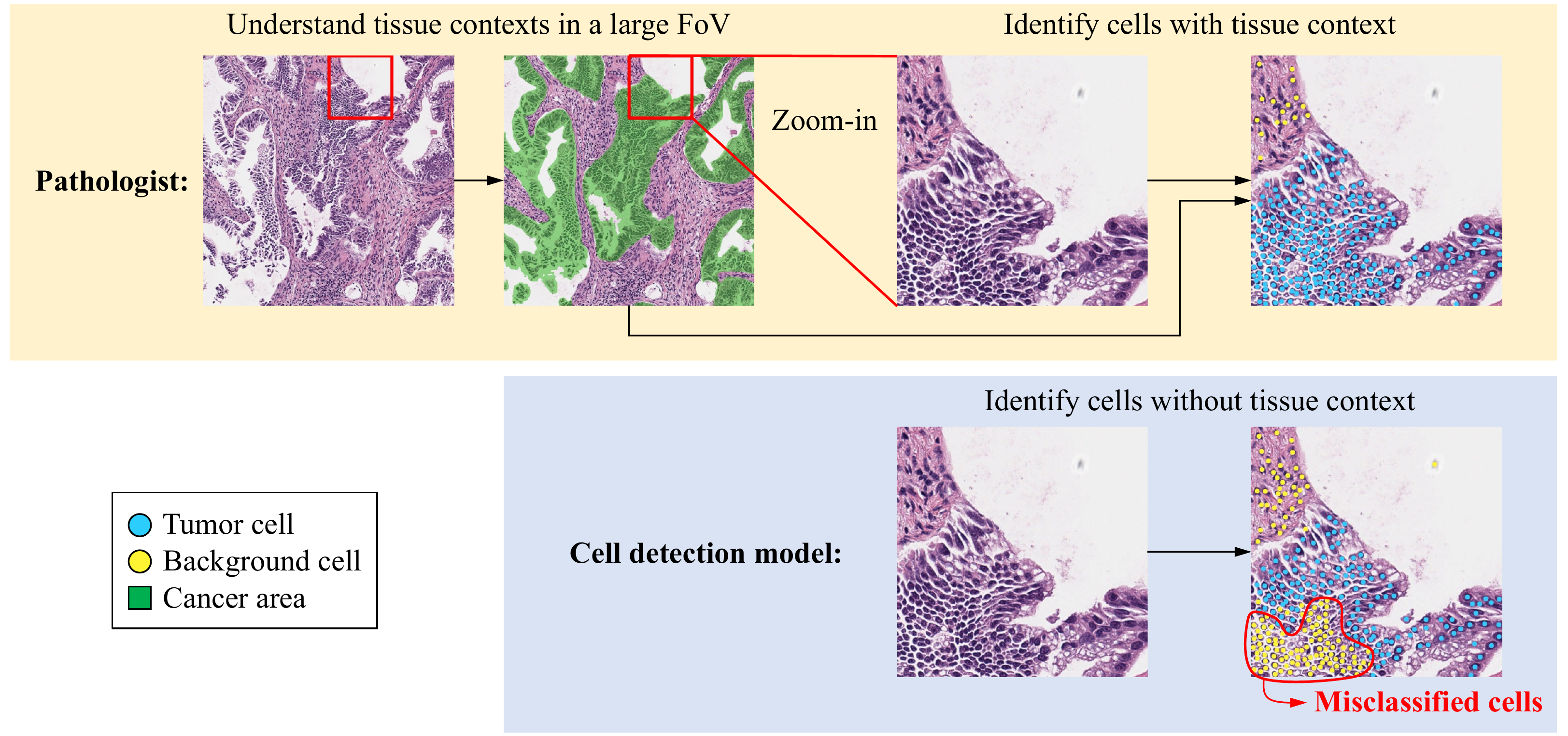}
    \hfill
    \vskip -3mm
    \caption{\textbf{Behavior of pathologists and cell detection models.} Typical cell detection models infer their predictions only by looking at a limited FoV region. Failure cases can occur when cells are difficult to be classified solely by their morphology, i.e., while disregarding the larger architecture of the tissues (context). In the above example, some groups of tumor cells are misclassified as background cells due to their morphology; these tumor cells are smaller and rounder than the nearby ones. Note that the large size and irregular shape are representative characteristics of tumor cells.
    Pathologists overcome these limitations by further understanding the tissue context of the larger FoV region. Misclassified cells can be easily corrected by understanding that such regions are cancer areas.
    }
    \label{fig:intro}
    \vskip -1mm
\end{figure*}

Computational Pathology (CPATH)~\cite{cpath} is a branch of digital pathology that develops methodologies for the analysis of digitized patient specimens, such as Whole-Slide-Images (WSIs).
Cell detection in histology images~\cite{ZHAO2020101786, lal2021nucleisegnet, qu2020weakly} is one of the most important tasks in CPATH. It allows the quantification and analysis of different cell types, which can lead to better prognosis evaluation~\cite{SUN2021103492,park2022artificial} and patient treatment planning while maintaining medical interpretability~\cite{diao2021human}. 
Since it has the potential to impact human lives, high-performance cell detection models are essential in real-world applications and need to be investigated.

To better locate and classify cells, detailed morphological characteristics such as color and shape are crucial. 
Consequently, cell detection datasets are typically collected at high magnification but small Field-of-View (FoV).
However, this can make the cell detection model overly rely on appearance details, without understanding the broader context~\cite{awmfovnet}. 
This context can help cell detection by providing information about how cells are arranged and grouped together to form high-level tissue structures.
In practice, expert annotators (pathologists) first zoom out to understand these broad tissue structures.
Next, they zoom in to better classify individual cells while taking into account the context information, as depicted in \autoref{fig:intro}.

The behavior of pathologists can be transferred to deep learning, for instance, through a multi-task strategy combining cell detection tasks at high magnification and tissue segmentation at low magnification. This type of approach would allow the model to share knowledge across different tasks and FoVs. However, to train such an approach, a combined dataset with cell-tissue overlapping regions is required; unfortunately, most existing datasets only target a single task, either cell detection~\cite{MoNuSeg, GRAHAM2019101563} or tissue segmentation~\cite{PANDA, digestpath}. 

In this paper, we introduce a new research direction: \emph{studying cell-tissue relationships for cell detection.}
First, we publish the \textbf{\ours} dataset, which contains cell and tissue annotations in small and large FoV patches, respectively, with overlapping regions. Additionally, the data is collected from WSIs of multiple organs. 
This can provide the necessary data for researchers to study cell-tissue relationships and their effect on cell detection.
Second, we introduce simple multi-task learning approaches for cell detection that can benefit from cell-tissue relationships and demonstrate their advantages over 3 different datasets. These approaches consistently show better cell detection performance compared to the cell-only baseline, i.e., a model trained only with small FoV patches with the corresponding cell annotations.
We hope that our proposed \ours dataset and methods will encourage the CPATH community to learn how to reflect cell-tissue relationships better to improve cell detection.

Our contributions are 3-fold and summarized as follows,
\begin{itemize}
    \item The first work that exploits cell-tissue relationships for better cell detection, to the best of our knowledge. 
\vspace{-1mm}
    \item We release \ours, a dataset with overlapping cell and tissue annotations based on Hematoxylin and Eosin (H\&E) stained WSIs of multiple organs. 
\vspace{-1mm}
    \item We introduce several approaches that boost cell detection performance via multi-task learning, and empirically confirm that these methods generalize well across different datasets and histological stainings.
\end{itemize}

\section{Related Work}
\label{sec:related_work}

\subsection{Datasets for Cell detection and Tissue segmentation Tasks}
\label{sec:ct_datasets}
In recent years, numerous datasets have been released for tackling cell detection. Some of those works only target a single organ \cite{TNBC, GRAHAM2019101563}, while others consider multiple ones \cite{CPM17, MoNuSeg, MoNuSAC, PanNuke, NuCLS}. The availability of these datasets enables the CPATH community to push forward the development and improvement of cell detection models \cite{GRAHAM2019101563, ZHAO2020101786, liu2019nuclei, lal2021nucleisegnet, li2019dual, qu2020weakly}. In addition, tissue segmentation datasets have also been proposed for prostate \cite{NIR2018167, 8853320, PANDA}, colorectal \cite{tissue_ds_1}, brain \cite{miccai2014}, and multiple organs \cite{ADP}. Some examples of tissue segmentation works can be found in \cite{tissue_ds_2, chan2019histosegnet, zhu2021multi, chen2016dcan, qian2022transformer, li2016gland}. 
The dataset in \cite{digestpath} is composed of a cell detection subset and a tissue segmentation subset. However, the subsets are annotated independently and from different patient groups, and, therefore, there are no overlaps between the cell and tissue data.
Because of the lack of overlapping data in the aforementioned datasets, it is difficult to build an end-to-end framework to learn cell-tissue relationships by jointly training on the cell and tissue tasks.

The recently released TIGER dataset~\cite{tiger} contains both cell and tissue annotations to study tumor-infiltrating lymphocytes \cite{SALGADO2015259} in H\&E breast cancer WSIs.
All the cell-annotated areas exist inside the tissue-annotated area, however, this work does not propose nor initiate any effort toward the integration of both cell and tissue tasks. 

\begin{table}[t]
\small
\centering
\setlength{\tabcolsep}{0.2em}
\begin{tabular}{lrrcl}
\toprule

\textbf{Dataset} & \multicolumn{1}{c}{\textbf{Tissue Area}} & \multicolumn{1}{c}{\textbf{\# Cell}} & \hspace{0.5em} & \multicolumn{1}{c}{\textbf{Organs}}\\ 
\midrule

\ours & \quad4.267$cm^2$ & \quad114.7K & \hspace{0.5em} &multiple\\

TIGER & \quad2.536$cm^2$ & \quad50.8K & \hspace{0.5em} & breast\\

\midrule

\end{tabular}
\vskip -2mm
\caption{\textbf{Dataset comparison} in terms of physical tissue annotated area and total cell counting per dataset.}
\label{tab:label_stats_new}
\vspace{-3mm}
\end{table}
To further promote the development of methods that leverage the cell-tissue relationship for the task of cell detection, we propose \ours, which is designed to capture the hierarchical relationship between cells and tissues, especially in the tumor environment. 
OCELOT contains roughly two times more cell and tissue annotations than TIGER (see \autoref{tab:label_stats_new}).
Additionally, the data was collected from multiple organs to enable the investigation of the generalizability of cell-tissue relationships over various cancer types.
In the end, we utilized both \ours and TIGER to reveal the power of cell-tissue relationship for cell detection in \autoref{sec:experiments}.

\subsection{Leveraging Large Field of View}
\label{sec:fov}
Some studies \cite{KAMNITSAS201761, awmfovnet, VANRIJTHOVEN2021101890, SCHMITZ2021101996, HO2021101866, 10.1007/978-3-030-59722-1_37} extract a large FoV region as an additional input to improve detection/segmentation performance on smaller FoV regions.
\cite{KAMNITSAS201761} proposes a dual pathway 3D CNN for brain lesion segmentation, where each pathway receives both small and large center-shared FoV regions as input. 
Similar studies are also conducted in the CPATH domain for tissue segmentation \cite{awmfovnet, VANRIJTHOVEN2021101890, SCHMITZ2021101996, HO2021101866} and cell detection \cite{10.1007/978-3-030-59722-1_37}.
To fuse different FoV patches, \cite{awmfovnet} introduces a weighting mechanism and \cite{VANRIJTHOVEN2021101890} proposes a multi-scale merging block composed of convolution and concatenation. 
Nevertheless, no inter-task relationship is considered in the previously mentioned methods. In contrast, our models take advantage of the large contextual information while learning the cell-tissue relationship via multi-task objectives at different FoVs.

\subsection{Leveraging Cell-Tissue Relationships}
\label{sec:ct_methods}
To our understanding, there is no study that considers cell-tissue relationships for cell detection or tissue segmentation tasks.
On the other hand, a few efforts have attempted to link tissue and cell for image classification using graph-based methods \cite{cgcnet, wsptcgcn, hactnet}.
Such studies represent the tissue structure as a graph of detected cells, based on the medical knowledge that cells form tissues.
\cite{hactnet} explicitly considers the cell-tissue relationship for the task of breast cancer subtyping, by an interaction between tissue-level and cell-level graphs with a cell-to-tissue hierarchy. 
However, these methods treat the cell/tissue graph generation as a pre-processing step, by using the inference output of independently pre-trained cell detection and tissue segmentation models. In contrast, we directly target the improvement of cell detection by utilizing the cell-tissue relationship. 

\section{\ours}
\label{sec:dataset}
\begin{figure}[t]
    \centering
    \hfill
        \includegraphics[width=\linewidth]{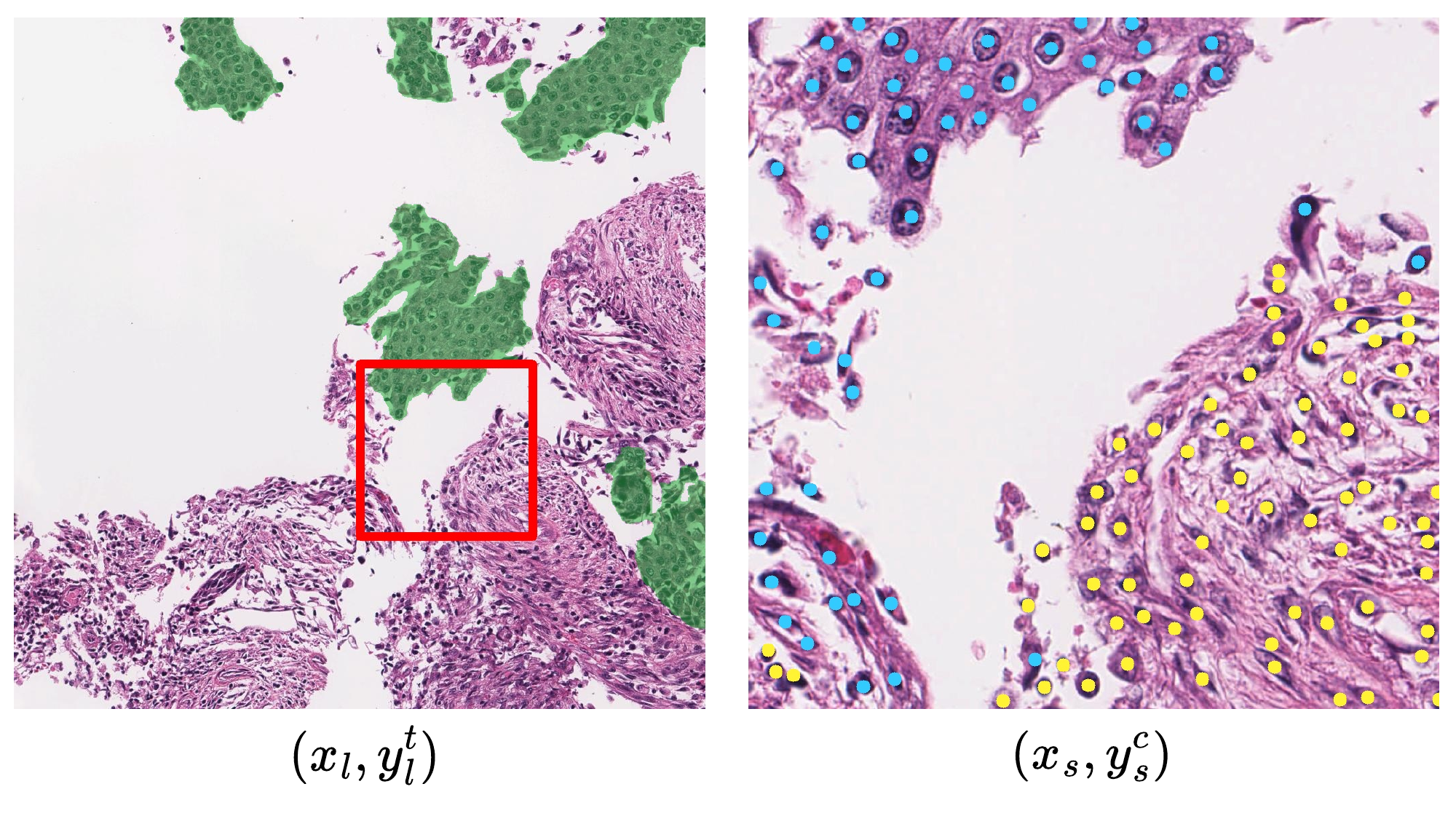}
    \hfill
    \vskip -7mm
    \caption{\textbf{A sample from the \ours dataset.} Each sample of the dataset consists of two input patches and the corresponding annotations. \textbf{Left} shows the large FoV patch $x_{l}$ with tissue segmentation annotation $y_{l}^{t}$, where green denotes the cancer area. \textbf{Right} shows the small FoV patch $x_{s}$ with cell point annotation $y_{s}^{c}$, where blue and yellow dots denote \textit{tumor} and \textit{background} cells, respectively. The red box indicates the size and location of the $x_{s}$ with respect to the $x_{l}$.}
    \label{fig:data_sample}
    \vspace{-0.3cm}
\end{figure}

In this section, we introduce \ours, a histopathology dataset specifically built to enable the development of methods that leverage cell and tissue relationships. Each sample of the \ours dataset $\mathcal{D}$ is composed of six components,

\begin{equation}
\mathcal{D} = \left\{\left(x_{s}, y_s^{c}, x_l, y_l^{t}, c_x, c_y\right)_i\right\}_{i=1}^{N}
\end{equation}

\noindent where $x_s, x_l$ are the small and large FoV patches extracted from the WSI, $y_s^{c}, y_l^{t}$ refer to the corresponding cell and tissue annotations, respectively, and $c_x, c_y$ are the relative coordinates of the center of $x_s$ within $x_l$. We drop the sample index $i$ for simplicity. \autoref{fig:data_sample} shows the visualization of a sample in \ours. More details about the dataset including data collection and statistics can be found in the following sub-sections. The dataset is publicly available at \href{https://lunit-io.github.io/research/publications/ocelot/}{https://lunit-io.github.io/research/publications/ocelot/}.

\subsection{Data Collection}
\label{ssec:data-collect}
We collect 306 TCGA~\cite{HUTTER2018283} WSIs from a total of 6 different organs: \textit{kidney}, \textit{head-and-neck}, \textit{prostate}, \textit{stomach}, \textit{endometrium}, and \textit{bladder}. 
From each of the WSIs, we select 1 to 3 large Regions of Interest (ROIs) for the tissue segmentation task. 
Finally, for the cell detection task, we randomly choose a smaller ROI that is fully contained within the larger tissue ROI. 
As a result, \ours includes 673 paired patches from 6 organs. The numbers of WSIs and pairs of patches per organ are detailed in \autoref{tab:dst_size}.

Some natural image datasets, such as ImageNet~\cite{deng2009imagenet} or Pascal VOC~\cite{everingham2010pascal}, include thousands of annotated images. However, annotating histopathology images is more challenging and expensive due to the scarcity of expert pathologists~\cite{c3det}. Furthermore, acquiring dense annotations for cell detection and tissue segmentation is especially time-demanding compared to higher-level tasks such as image classification. Nonetheless, in \autoref{tab:label_stats_new}, we observe that \ours is roughly double the size of the recent TIGER dataset with respect to the annotated tissue area and the number of annotated cells.

\vspace{-4mm}
\paragraph{Patch configuration.} 
Cell detection tasks benefit from fine-grained spatial information to better capture detailed cell properties (e.g. border, shape, color, and opacity). In contrast, tissue segmentation requires a larger context to enable a better understanding of the overall structural information. Therefore, we define the FoV sizes of $x_s$ (cell detection) and $x_l$ (tissue segmentation) as 1024 $\times$ 1024 and 4096 $\times$ 4096 pixels, respectively, at a resolution of 0.2 Microns-per-Pixel (MPP). Finally, the large FoV patches and tissue annotations ($x_l$, $y_l^{t}$) are down-sampled by a factor of 4, resulting in a size of 1024 $\times$ 1024 pixels.

\vspace{-4mm}
\paragraph{Annotation.} 
All cell-tissue pairs of patches are annotated by board-certified pathologists. Cells are labeled as points, with associated 2D coordinates and class labels. We denote the annotations in a given cell-level patch, $x_s$, as $y_s^{c}$, and consider two classes: Tumor Cell (\textit{TC}) and Background Cell (\textit{BC})\footnote{\textit{BC} includes any of the following cell categories: lymphocyte, macrophage, fibroblast, endothelial, or other remaining cell types.}.
\textit{TC} and \textit{BC} class ratios are 35.01\% and 64.99\%, respectively.
Regarding the tissue patches, $x_l$, pathologists annotate the pixel-wise segmentation maps $y_l^{t}$ with either Cancer Area (\textit{CA}) or Background (\textit{BG}) labels. A minority of pixels where the tissue class was uncertain were labeled as Unknown (\textit{UNK}). \textit{BG}, \textit{CA}, and \textit{UNK} class ratios are 55.77\%,  40.17\%, and 4.06\%, respectively. The amount of annotated cells and tissue pixels, per data split, can be found in the \supple. The detection of \textit{TC} and \textit{BC} has clinical relevance. For example, tumor purity \cite{azimi2017breast}, computed as the tumor/non-tumor cell ratio in a WSI, has a correlation with cancer prognosis \cite{mao2018low, zhang2017tumor, gong2020tumor}.

\vspace{-4mm}
\paragraph{Dataset splits.} The dataset is divided into three subsets: \textit{training}, \textit{validation}, and \textit{test}, following a $6$:$2$:$2$ ratio. To prevent information leaking among the data subsets, we randomly split the dataset per WSI, so that different patches from the same WSI are not included in multiple subsets. We maintain consistent cancer-type ratios in each subset.

\begin{table}
\small

\centering
\setlength{\tabcolsep}{0.2em}
{
\begin{tabular}{lrrrrrr}
\toprule
\multirow{2}{*}{\textbf{Organs}} & \multicolumn{3}{c}{\textbf{\# Slides}} & \multicolumn{3}{c}{\textbf{\# Patch Pairs}} \\ \cmidrule(lr){5-7} \cmidrule(lr){2-4} 
& \textbf{Train} & \textbf{Val} & \textbf{Test} &  \textbf{Train} & \textbf{Val} & \textbf{Test} \\ \midrule
Kidney & 48 & 15 & 18 & 125 & 41 & 41 \\ 
Head-neck & 13 & 5 & 6 & 27 & 9 & 10 \\ 
Prostate & 26 & 12 & 10 & 50 & 17 & 16 \\ 
Stomach & 15 & 6 & 5 & 36 & 12 & 12 \\ 
Endometrium & 38 & 13 & 13 & 86 & 29 & 25 \\ 
Bladder & 35 & 14 & 14 & 82 & 29 & 26 \\ \midrule
\textbf{Total} & 175 & 65 & 66 & 406 & 137 & 130 \\ \bottomrule
\end{tabular}
}
\vskip -2mm
\caption{\textbf{Dataset size per organ and data subset.}}
\label{tab:dst_size}
\vspace{-4mm}
\end{table}
\section{Empirical Analysis}
\label{sec:motivation}
The motivation for considering the cell-tissue relationships for the development of cell detection models stem from the biological and hierarchical arrangement of cells and tissues. These insights are further corroborated by two main empirical observations described in the following \autoref{sec-class-corr} and \autoref{poc-exp}.

\subsection{Interrelation between cell and tissue classes} \label{sec-class-corr}
\begin{table}[t]
\small
\addtolength{\leftskip} {-2cm}
\addtolength{\rightskip}{-2cm}
\setlength{\tabcolsep}{0.3em}
\centering
{
\begin{tabular}{lrrclrr}
\cmidrule[\heavyrulewidth]{1-3}\cmidrule[\heavyrulewidth]{5-7}
\multirow{2}[2]{*}{\textbf{Cell}}   & \multicolumn{2}{c}{\textbf{Tissue}}                                       & \hspace{2em} & \multirow{2}[2]{*}{\textbf{Cell}} & \multicolumn{2}{c}{\textbf{Tissue}} \\ \cmidrule(l){2-3}\cmidrule(l){6-7}
                                    & \multicolumn{1}{c}{\textit{CA}}   & \multicolumn{1}{c}{\textit{non-CA}}   &  &                                   & \multicolumn{1}{c}{\textit{ST}} & \multicolumn{1}{c}{\textit{non-ST}} \\ \cmidrule{1-3}\cmidrule{5-7}
\textit{TC}\enspace                 & 67.7K             & 5.4K          & & \multirow{2}{*}{\textit{LC}\enspace}    & \multirow{2}{*}{45.4K}   & \multirow{2}{*}{5.4K\enspace} \\
\textit{BC}                         & 6.4K              & 35.2K         & &                                         &                                   &                               \\ \cmidrule[\heavyrulewidth]{1-3}\cmidrule[\heavyrulewidth]{5-7}
\multicolumn{3}{c}{(a) \ours} && \multicolumn{3}{c}{(b) TIGER} \\
\end{tabular}
}
\caption{\textbf{Cell counts based on the tissue class.}
Each value stands for the number of cells located inside the tissue area. \textit{TC}, \textit{BC}, \textit{LC}, \textit{CA}, and \textit{ST} stand for Tumor Cell, Background Cell, Lymphocyte Cell, Cancer Area tissue, and Stroma tissue, respectively. 
}
\label{tab:stat}
\end{table}

\begin{table}[t]
\addtolength{\leftskip} {-2cm}
\addtolength{\rightskip}{-2cm}
\setlength{\tabcolsep}{0.3em}
\centering
{
\begin{tabular}{rrcrr}
\cmidrule[\heavyrulewidth]{1-2}\cmidrule[\heavyrulewidth]{4-5}
\multicolumn{1}{c}{\textit{CA}} & \multicolumn{1}{c}{\textit{non-CA}} &\hspace{2em} & \multicolumn{1}{c}{\textit{ST}} & \multicolumn{1}{c}{\textit{non-ST}} \\
\cmidrule{1-2}\cmidrule{4-5}
40.17\% & \enspace59.83\% && 30.79\% & \enspace69.21\% \\
\cmidrule[\heavyrulewidth]{1-2}\cmidrule[\heavyrulewidth]{4-5} 
\multicolumn{2}{c}{(a) \ours} && \multicolumn{2}{c}{(b) TIGER} \\
\end{tabular}
}
\caption{\textbf{Pixel ratio among tissue classes.} 
\textit{CA}, and \textit{ST} stand for Cancer Area tissue, and Stroma tissue, respectively.}
\label{tab:stat2}

\vspace{-4mm}
\end{table}

We empirically observe the interrelation between specific cell and tissue classes by counting the amount of each annotated cell type within each tissue region, as observed in \autoref{tab:stat}. Indeed, we verify that in \ours, around 93\% of \textit{TC} are located within \textit{CA} and 85\% of \textit{BC} are found outside of the \textit{CA} tissue. Note that \textit{CA} is not the majority tissue class (\autoref{tab:stat2}), therefore, we conclude that there is, in fact, a relationship between the cell and tissue classes. We observe a similar phenomenon in the TIGER dataset when considering the \textit{LC} and \textit{ST} classes (\autoref{tab:stat} and \autoref{tab:stat2}).

In practice, pathologists classify cells by taking into account such interrelationships, since it is difficult to classify isolated cells, without considering the larger context of the tissues. 
As depicted in \autoref{fig:intro}, pathologists first need to visualize the structure of the tissues at a larger FoV. Then, they zoom in and consider the previously observed context along with the fine-grained details of each individual cell and nearby neighborhood, thus considering the cell-tissue dependencies. 
Inspired by the behavior of pathologists, we expect a cell detection model to also benefit from understanding the tissue structure from a broader viewpoint.

\subsection{Tissue-label Leaking Model} \label{poc-exp}
\begin{figure}[t]
    \centering
    \includegraphics[width=0.9\linewidth]{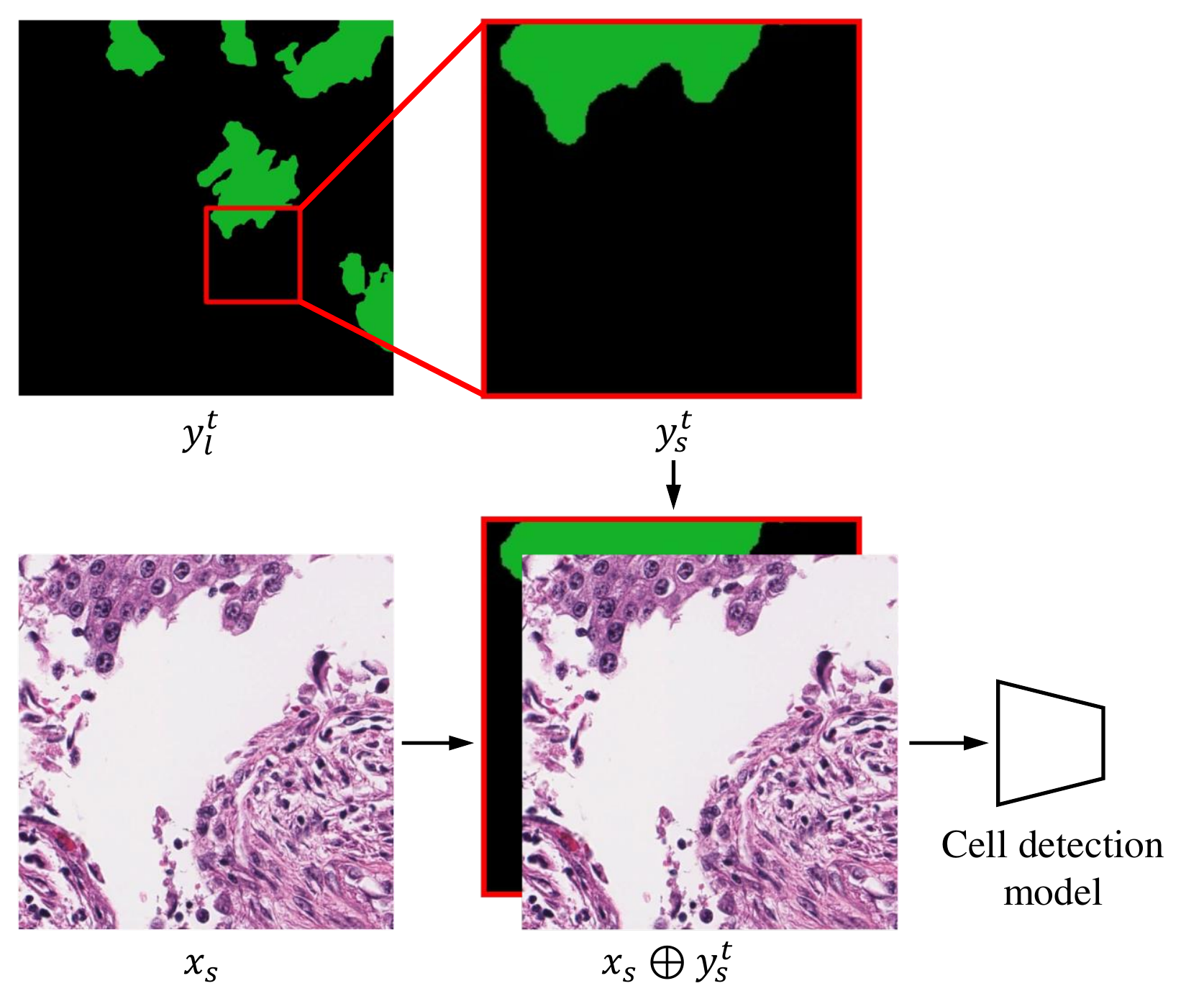}
    \caption{  
        \textbf{Tissue label leaking model.} 
        This model receives the cell patch $x_s$ along with the corresponding tissue labels as input. 
        The region corresponding to the cell patch in the tissue patch annotation is cropped, upsampled, and concatenated to the cell patch. $\oplus$ denotes channel-wise concatenation.
    }
    \label{fig:poc}
    \vspace{-0.3cm}
\end{figure}

In the previous section, we observe a strong relationship between certain cell and tissue classes. To further validate the hypothesis that a cell detection model can leverage information from the tissue structure, we design an exploratory experiment where the tissue annotation is provided as an extra input to a cell detection model. To this end, we first crop the corresponding cell patch region from the tissue annotation $y_l^t$, and upsample it to match the size of the cell patch, $x_s$; the cropped tissue annotation is denoted as $y_s^t$. Finally, we concatenate $x_s$ and $y_s^t$ at the channels dimension and use this data to train a cell detection model. We denote this model as a \textit{Tissue-label leaking} model and illustrate it in \autoref{fig:poc}. Note that this model is not appropriate for real-world scenarios as the tissue labels are unknown at inference time, and is explored for the purpose of empirical analysis.

When we compare the performance of the \textit{tissue-label leaking} model with the standard cell detection model on the \ours dataset, we observe a significant improvement in terms of mean F1-score\footnote{True positive (TP), false positive (FP), and false negative (FN) counts are determined following \cite{SwiderskaChadaj2019LearningTD}. If a detected cell is within a valid distance ($\approx 3\mu m$) from an annotated cell and the cell class matches, it is counted as a TP, otherwise an FP. If an annotated cell is not detected, it is counted as an FN. Then, the mean F1 score across classes is computed.} performance of \textbf{+7.69} and \textbf{+9.76} in the validation and test sets, respectively.
Detailed results can be found in \autoref{tab:main}. Taking these results into consideration, we conclude that there is significant room to improve the cell detection model, which can be achieved by combining the tasks of cell detection and tissue segmentation. 

\section{Method}
\label{sec:approach}
In this section, we propose to utilize cell-tissue relationships through multi-task learning. 
First, we propose a set of approaches inspired by the \textit{tissue-label leaking} model described in \autoref{poc-exp}. These models replace the annotated tissue labels with predictions from an auxiliary tissue segmentation branch.
Second, we design a bi-directional information-sharing approach that shares features in both tissue-to-cell and cell-to-tissue directions. The proposed approaches are described in \autoref{subsec:pred-to-x} and \autoref{subsec:arch-search}

\subsection{Preliminary} \label{subsec:preliminary}
To deal with cell-tissue sample pairs, i.e., $(x_s, y_s^c)$ and $(x_l, y_l^t)$, we build a dual-branch architecture containing separate networks for the cell and tissue tasks. Similarly to \cite{SwiderskaChadaj2019LearningTD}, we define cell detection as a segmentation task. Specifically, the cell labels are provided as a segmentation map by drawing a fixed-radius circle centered on each cell point annotation and filled with the corresponding class label.
At inference time, we find local peaks within the predicted cell probability maps and output them as point predictions.
More details are provided in the supplementary material.
Treating cell detection as a segmentation task enables us to use the same architecture for both cell and tissue branches, which largely simplifies the training and tuning of the model and reduces the range of possible decisions, such as neural network architecture, or hyper-parameters.
We use DeepLabV3+~\cite{deeplabv3plus2018} as a base architecture for both branches and single-task models. 

\subsection{Tissue-prediction Injection Models} \label{subsec:pred-to-x}
\begin{figure}[t]
	\centering
    \includegraphics[width=\linewidth]{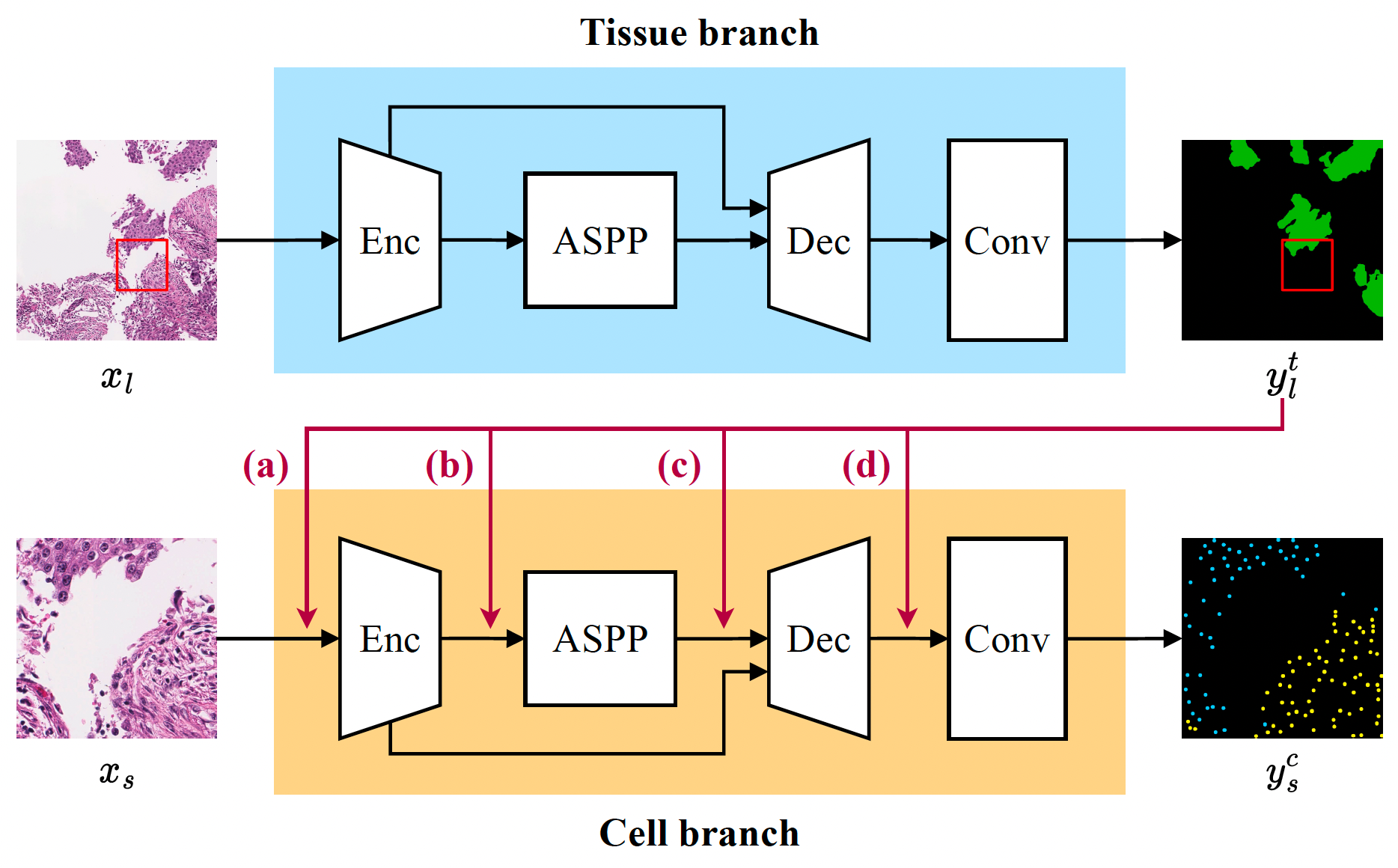}
	\caption{\textbf{Tissue-prediction injection model} injects the tissue segmentation prediction into 1 out of 4 locations of the cell branch: (a) input, (b) after encoder, (c) after ASPP, and (d) after decoder.}
	\label{fig:pred_to_x}
    \vspace{-0.3cm}
\end{figure}

These models are a simple and practical extension of the \textit{tissue-label leaking} model, where we inject the predicted tissue probabilities into the cell detection branch instead of leaking the tissue labels. We consider only one injection point in the cell detection branch, but explore four possible alternatives: (a) at the input (\textit{Pred-to-input}), (b) after the encoder (\textit{Pred-to-inter-1}), (c) after the ASPP module (\textit{Pred-to-inter-2}), and (d) after the decoder (\textit{Pred-to-output}). \autoref{fig:pred_to_x} depicts this family of models, denominated as \textit{Tissue-prediction injection}.
Since the tissue and cell patches represent different regions, we need to align the content between the tissue and cell feature maps before concatenation. Therefore, we crop the cell corresponding region from the tissue predictions, upsample them, and, finally, concatenate them in the channel dimensions of the feature maps of the cell branch, as illustrated in \autoref{fig:tissue_to_cell}.

\subsection{Cell-Tissue Feature Sharing Model} \label{subsec:arch-search}
\begin{figure}[t]
    \centering
    \hfill
    \begin{subfigure}{0.44\linewidth}
        \includegraphics[width=\linewidth]{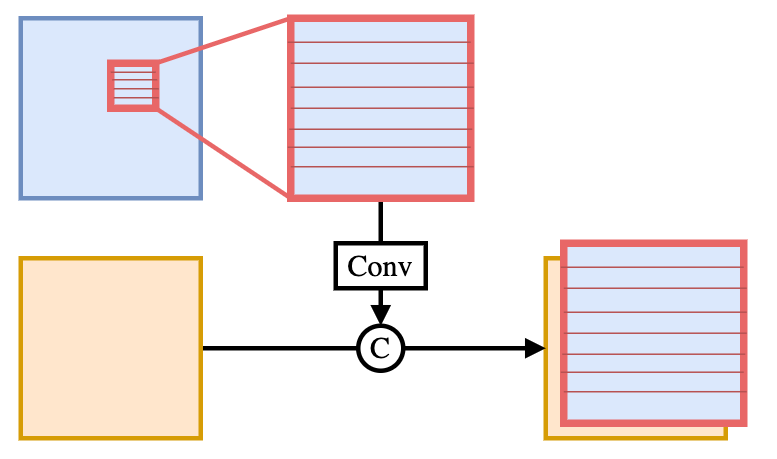}
        \caption{Tissue to Cell}
        \label{fig:tissue_to_cell}
        \end{subfigure}
    \hfill
    \begin{subfigure}{0.48\linewidth}
        \includegraphics[width=\linewidth]{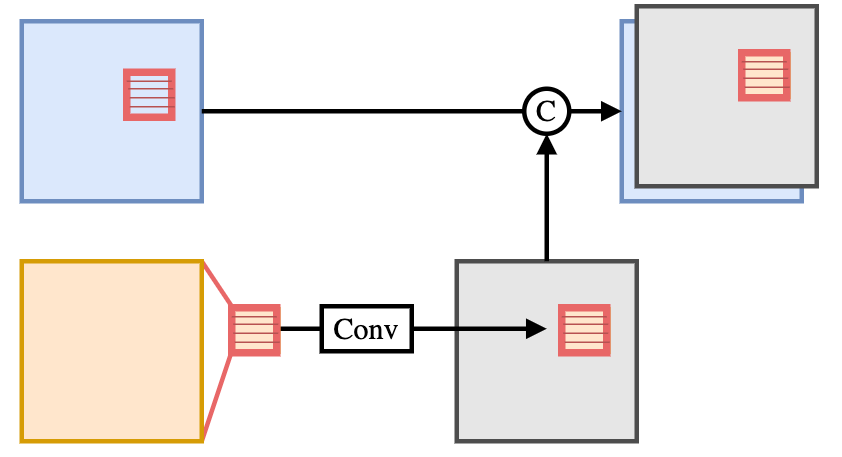}
        \caption{Cell to Tissue}
        \label{fig:cell_to_tissue}
    \end{subfigure}
    \hfill
	\caption{Information is shared between cell and tissue branches via channel-wise concatenation preceded by a shallow convolutional layer with $3\times3$ kernel size. Cropping and upsampling (in \autoref{fig:tissue_to_cell}) or downsampling and zero-padding (in \autoref{fig:cell_to_tissue}) is applied to match the patch sizes and pixel-alignment between two feature maps from different FoVs. The cell and tissue feature maps are represented in orange and blue, respectively. The red contour denotes the cell patch-associated region in the tissue patch, and the gray regions represent zero padding.}
    \label{fig:info_exchange}
\end{figure}

\begin{figure}[t]
	\centering
    \includegraphics[width=\linewidth]{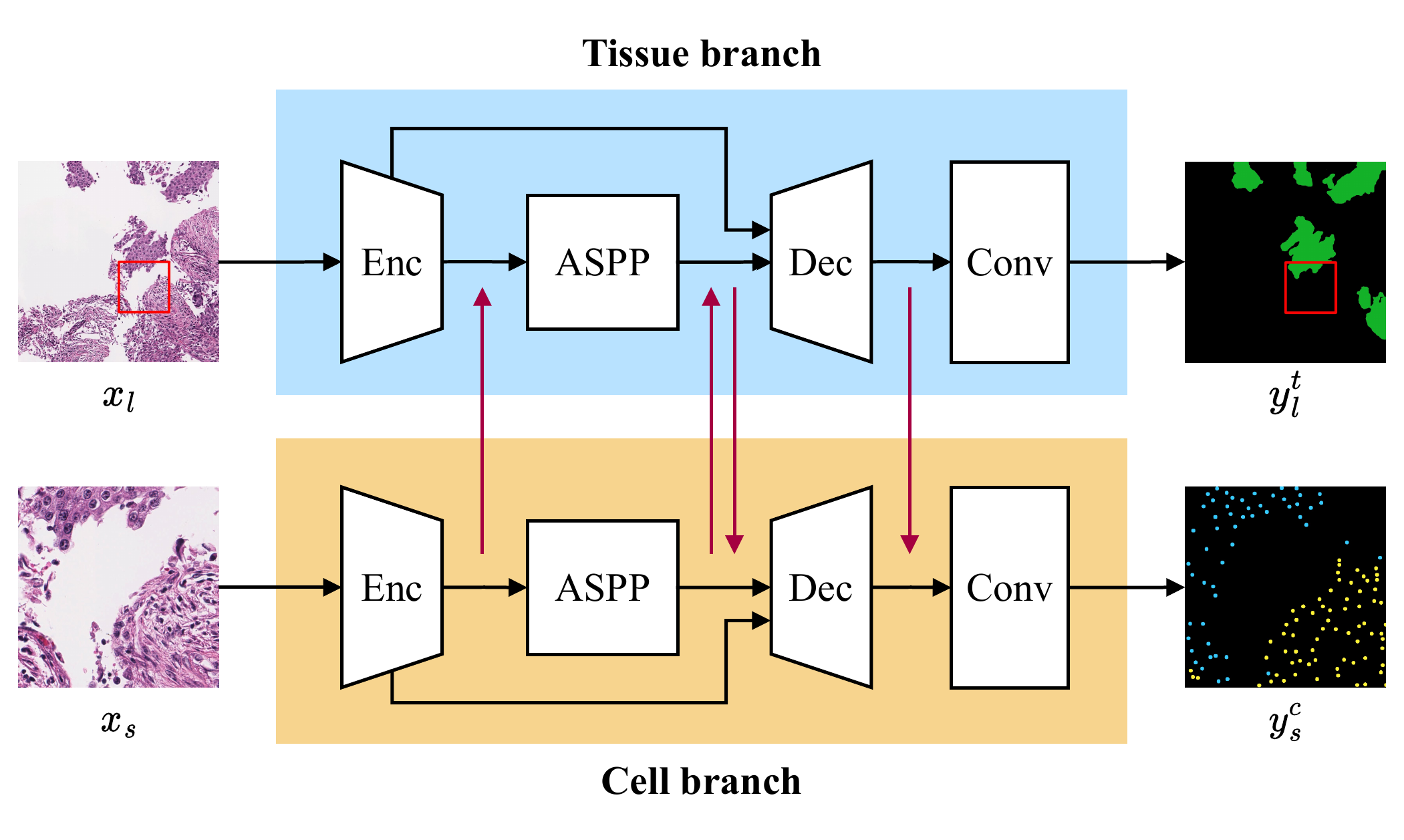}
	\caption{\textbf{Cell-Tissue Feature Sharing Model} has two branches for tissue segmentation and cell detection.  Information exchange occurs multiple times between the two branches, indicated by the red vertical arrows. Details regarding the information exchange procedures are described in \autoref{fig:info_exchange}.}
	\label{fig:our_model}
\vspace{-4mm}
\end{figure}

\textit{Tissue-prediction injection} models share the tissue prediction in a single location and direction, i.e., tissue-to-cell. To enable a more diverse and flexible cell-tissue information flow, we also explore bi-directional feature map sharing from cell-to-tissue (\autoref{fig:info_exchange}, left) and tissue-to-cell (\autoref{fig:info_exchange}, right). Considering these two operations, we conduct an architecture search procedure to find the optimal feature map sharing configuration between both branches. To limit the search space, we constrain it to only 3 positions in the architecture: after the encoder, after the ASPP module, or after the decoder. Furthermore, we also exclusively allow the branches to inject feature maps at the same depth or position.
Finally, we consider only the best-performing model among the $4^3$ candidates, which is presented in \autoref{fig:our_model}. 
We name this model as \textit{cell-tissue feature sharing}.

\section{Experiments and Results}
\label{sec:experiments}
\begin{table*}[t]
\addtolength{\leftskip} {-2cm}
\addtolength{\rightskip}{-2cm}
\setlength{\tabcolsep}{0.3em}
\centering
{
  \begin{tabular}{lrrrrrr}
    \toprule
    \multicolumn{1}{c}{\multirow{2}{*}{\textbf{Method}}} & \multicolumn{2}{c}{\textbf{\ours}} & \multicolumn{2}{c}{\textbf{TIGER}} & \multicolumn{2}{c}{\textbf{CARP}} \\ \cmidrule(lr){2-3}\cmidrule(lr){4-5}\cmidrule(lr){6-7} 
                    & \multicolumn{1}{c}{Val} & \multicolumn{1}{c}{Test} & \multicolumn{1}{c}{Val} & \multicolumn{1}{c}{Test} & \multicolumn{1}{c}{Val} & \multicolumn{1}{c}{Test} \\ \midrule
    \textit{Cell-only}       & \score{68.87}{1.76}   & \score{64.44}{1.82}   & \score{63.89}{1.39}   & \score{53.82}{1.23}   & \score{78.48}{0.69}   & \score{70.96}{1.47}   \\
    \textit{Pred-to-input}   & \textbf{\score{73.36}{0.59}} & \score{69.65}{3.93} & \underline{\score{66.00}{2.00}} & \score{53.29}{1.30} & \score{79.46}{0.79} & \score{72.98}{0.82} \\ 
    \textit{Pred-to-inter-1} & \underline{\score{72.74}{0.50}} & \underline{\score{70.54}{2.20}} & \textbf{\score{66.19}{1.02}} & \textbf{\score{55.87}{1.78}} & \underline{\score{79.74}{0.80}} & \underline{\score{73.05}{0.69}} \\
    \textit{Pred-to-inter-2} & \score{72.68}{1.58}   & \textbf{\score{71.23}{0.96}} & \score{65.43}{1.14} & \underline{\score{54.75}{2.25}} & \textbf{\score{79.87}{0.78}} & \textbf{\score{73.14}{1.53}} \\
    \textit{Pred-to-output}  & \score{66.85}{5.62}   & \score{65.05}{3.72}   & \score{63.02}{0.16}   & \score{53.32}{0.42}   & \score{78.92}{0.60}   & \score{72.61}{0.95}   \\
    \textit{Feature-sharing} & \score{72.30}{0.73}   & \score{68.91}{2.52}   & \score{65.64}{1.07}   & \score{55.10}{2.18}   & \score{79.38}{0.74}   & \score{73.00}{0.33}   \\ \midrule
    \textit{Tissue-label leaking}\enspace   & \score{76.56}{0.80}   & \score{74.20}{0.91}   & \score{69.71}{0.61}   & \score{61.66}{1.16}   & \score{80.13}{1.04}   & \score{72.97}{0.49}   \\ \bottomrule
  \end{tabular}
}
\caption{\textbf{Cell detection mean F1 scores per model.} With the exception of the \textit{Tissue-label Leaking} model, the \textbf{highest score} is written in bold and the \underline{second highest score} is underlined.
\label{tab:main}
}
\end{table*}
\begin{table*}[t]
\addtolength{\leftskip} {-2cm}
\addtolength{\rightskip}{-2cm}
\setlength{\tabcolsep}{0.3em}
\centering
{
  \begin{tabular}{lccccccc}
    \toprule
    \multicolumn{1}{c}{\multirow{2}{*}{\textbf{Method}}} &\multicolumn{1}{c}{population} & \multicolumn{1}{c}{kidney} & \multicolumn{1}{c}{endometrium} & \multicolumn{1}{c}{bladder} & \multicolumn{1}{c}{prostate} & \multicolumn{1}{c}{stomach} & \multicolumn{1}{c}{head-neck} \\ 
    &
    {\footnotesize(130)}&\multicolumn{1}{c}
    {\footnotesize(41)}&\multicolumn{1}{c}{\footnotesize(25)}&\multicolumn{1}{c}{\footnotesize(26)}&\multicolumn{1}{c}{\footnotesize(16)}&\multicolumn{1}{c}{\footnotesize(12)}&\multicolumn{1}{c}{\footnotesize(10)}\\ \midrule 
    \textit{Cell-only} & \score{64.44}{1.82} & \score{64.12}{3.41} & \score{66.88}{3.71} & \score{59.42}{5.61} & \score{65.46}{1.64} & \score{66.19}{8.93} & \score{59.24}{5.04} \\
    \textit{Pred-to-input} & \score{69.65}{3.93} & \score{63.37}{3.27} & \underline{\score{73.31}{6.35}} & \score{63.61}{4.55} & \underline{\score{68.36}{8.61}} & \score{69.34}{7.15} & \textbf{\score{75.28}{0.82}} \\ 
    \textit{Pred-to-inter-1} & \underline{\score{70.54}{2.20}} & \underline{\score{66.62}{7.30}} & \score{73.05}{4.14} & \underline{\score{64.35}{3.59}} & \textbf{\score{70.20}{2.56}} & \textbf{\score{71.55}{5.21}} & \underline{\score{74.50}{1.52}} \\
    \textit{Pred-to-inter-2} & \textbf{\score{71.23}{0.96}} & \textbf{\score{68.94}{5.65}} & \textbf{\score{75.15}{2.70}} & \textbf{\score{64.94}{2.34}} & \score{68.35}{8.83} & \underline{\score{70.29}{0.81}} & \score{73.74}{1.72} \\
    \textit{Pred-to-output} & \score{65.05}{3.72} & \score{63.38}{4.48} & \score{68.21}{4.11} & \score{59.81}{3.96} & \score{64.18}{8.41} & \score{67.51}{6.39} & \score{60.88}{6.77} \\
    \textit{Feature-sharing} & \score{68.91}{2.52} & \score{64.97}{5.33} & \score{71.28}{5.30} & \score{63.21}{5.31} & \score{66.25}{9.52} & \score{69.79}{2.25} & \score{73.88}{4.39} \\ \midrule
    \textit{Tissue-label leaking}\enspace & \score{74.20}{0.91} & \score{75.75}{1.97} & \score{72.71}{1.12} & \score{71.18}{0.93} & \score{74.56}{1.25} & \score{75.24}{1.71} & \score{75.78}{0.46} \\ \bottomrule
  \end{tabular}

\caption{\textbf{Per-organ cell detection mean F1 scores in the \ours test set.} 
Population mean F1 scores are computed from all the patches in the test set.
In parentheses, we indicate the number of samples of each organ subset. With the exception of the \textit{Tissue-label Leaking} model, the \textbf{highest score} is written in bold and the \underline{second highest score} is underlined.}
\label{tab:organ}
\vspace{-4mm}
}
\end{table*}
We validate the hypothesis that incorporating cell-tissue relationships within a cell detection model is beneficial by evaluating the proposed models on \ours and 2 other datasets. 
First, we describe the additional datasets and implementation details. 
Then, we show how the proposed multi-task learning methods can improve the performance of a cell detection task. 
Lastly, we conduct an ablation study to investigate how important using large FoV patches $\{x_l\}$ and corresponding tissue segmentation labels $\{y_l^t\}$ are in enhancing cell-tissue relationships.

\subsection{Other Datasets}
\label{ssec:other-datasets}
\paragraph{TIGER.} 
As mentioned in \autoref{sec:ct_datasets}, TIGER is a public dataset, based on H\&E stained images, which includes both cell and tissue annotations in overlapping patches.
However, the patch sizes and amount of overlap between cell and tissue annotations highly vary among samples. Therefore, a pre-processing step is necessary to generate paired cell-tissue patches with consistent FoVs. 
After pre-processing, we obtain 9,888 paired patches of size 512$\times$512 and 128$\times$128 pixels for the tissue segmentation and cell detection tasks, respectively, with an MPP of approximately 0.5 $\mu m/px$.
Most samples in TIGER include small regions with annotated cells, making it necessary to extract smaller patches. This explains why the absolute number of patches is larger than in \ours or CARP, although the actual amount of annotated cells and tissue area are smaller than the latter datasets, as shown in \autoref{tab:label_stats_new}.
In TIGER, one cell class (lymphocyte cells) and seven tissue classes are annotated. More details about the pre-processing and this dataset are provided in the \supple.

\vspace{-4mm}
\paragraph{CARP.} This is an internal lung cancer dataset of PD-L1 IHC-stained WSIs, containing 6,480 paired patches extracted from 1,012 WSIs. The patch sizes, resolution, annotation protocol, and general configuration are similar to \ours, as described in \autoref{ssec:data-collect}.
Two cell classes are annotated: PD-L1 positive tumor cells (\textit{TC+}), and PD-L1 negative tumor cells (\textit{TC-}). The tissue classes are the same as in \ours (BG, CA, and UNK).
CARP is a real-world and large-scale dataset, with 809.1K annotated cells and 4.1080 $mm^2$ tissue area, which is approximately 10 and 20 times more than \ours and TIGER, respectively.
Moreover, the stain type is different from \ours and TIGER, allowing us to validate the effectiveness of the cell-tissue relationships across different staining methods.

\subsection{Implementation details}
As previously mentioned in \autoref{subsec:preliminary}, both cell and tissue branches are based on DeepLabV3+ \cite{deeplabv3plus2018}, with a ResNet-34~\cite{He2015} encoder. The models are trained for 300, 150, and 100 epochs in \ours, TIGER, and CARP datasets, respectively. We use the Adam optimizer~\cite{KingmaB14}, and, for each experiment, we tune the learning rate within a $[5\times10^{-5}, 2\times10^{-3}]$ range.
The model at the epoch with the best validation set performance is chosen and used for evaluation on the test set.
All experiments are repeated 5 times, and we report the mean and 95\% confidence interval of the performance metrics.
For more implementation details, please refer to the \supple.

\subsection{Main Results}
The cell detection results obtained by the \textit{Cell-only} baseline, the proposed approaches, and the \textit{Tissue-label leaking} model are shown in \autoref{tab:main}.
The \textit{Cell-only} baseline is a simple cell detection model that only receives the small FoV patches as input. In other words, it only considers the cell branch in \autoref{fig:pred_to_x} and neither tissue annotations nor large FoV patches are leveraged. The \textit{Tissue-label leaking} model, described in \autoref{subsec:pred-to-x}, receives the tissue annotations as input and serves as an exploratory experiment to obtain insight regarding how much the cell detection task can benefit from leveraging tissue annotation.

From \autoref{tab:main}, we observe that all the cell-tissue multi-task learning-based approaches, except for the \textit{Pred-to-output} model, outperform the \textit{Cell-only} baseline across all datasets. These improvements imply that cell detection on small FoV patches benefits from learning cell-tissue interrelationships from both large FoVs and tissue annotations.
We hypothesize that the reason for the low performance of the \textit{Pred-to-output} model is that the injection of tissue prediction to the cell branch happens too late. Therefore, there is a lack of network capacity for fusing cell and tissue information.

Furthermore, in \autoref{tab:organ}, we report mean F1 scores per organ in the  \ours test set. Except for the \textit{Pred-to-output}, all approaches show improvement compared to \textit{Cell-only} baseline in most organs. Especially, our best-performing model, \textit{Pred-to-inter-2}, demonstrates superior performance over \textit{Cell-only} in all 6 organs with a significant margin. Such a result shows that considering cell-tissue relationships can generally help the cell detection task across various organs.

\begin{table}[t]
{
\addtolength{\leftskip} {-2cm}
\addtolength{\rightskip}{-2cm}
\setlength{\tabcolsep}{0.3em}
\centering
\resizebox{\linewidth}{!}
{
    \begin{tabular}{ccrrrr}
    \toprule
    \multicolumn{2}{c}{Further utilize} & \multicolumn{2}{c}{\textit{Pred-to-inter-2}} & \multicolumn{2}{c}{\textit{Feature-sharing}} \\ \cmidrule(lr){3-4}\cmidrule(lr){5-6}
    Tissue label    & Large FoV     & \multicolumn{1}{c}{Val}   & \multicolumn{1}{c}{Test}  & \multicolumn{1}{c}{Val}   & \multicolumn{1}{c}{Test}  \\ \midrule
    -               & -             & \score{68.87}{1.76}       & \score{64.45}{1.82}       & \score{68.87}{1.76}       & \score{64.45}{1.82}  \\
    \checkmark      & -             & \score{70.21}{1.49}       & \score{64.99}{1.77}       & \score{70.36}{1.34}       & \score{66.58}{2.02} \\
    -               & \checkmark    & \score{68.78}{0.94}       & \score{65.56}{3.66}                & \score{69.61}{1.59}       & \score{66.22}{1.53} \\
    \checkmark      & \checkmark    &{\textbf{\score{72.68}{1.58}}} & \textbf{\score{71.23}{0.96}} & \textbf{\score{72.30}{0.73}} & \textbf{\score{68.91}{2.52}} \\
    \bottomrule
  \end{tabular}
}
\caption{\textbf{Dataset ablation study.} In all experiments, the model architecture is fixed and only input/output changes. The first row corresponds to the \textit{Cell-only} baseline. 
}
\label{tab:ablation}
\vspace{-6mm}
}
\end{table}

\begin{figure*}[t]
    \centering
    \vspace*{-2mm}
    \includegraphics[width=0.94\linewidth]{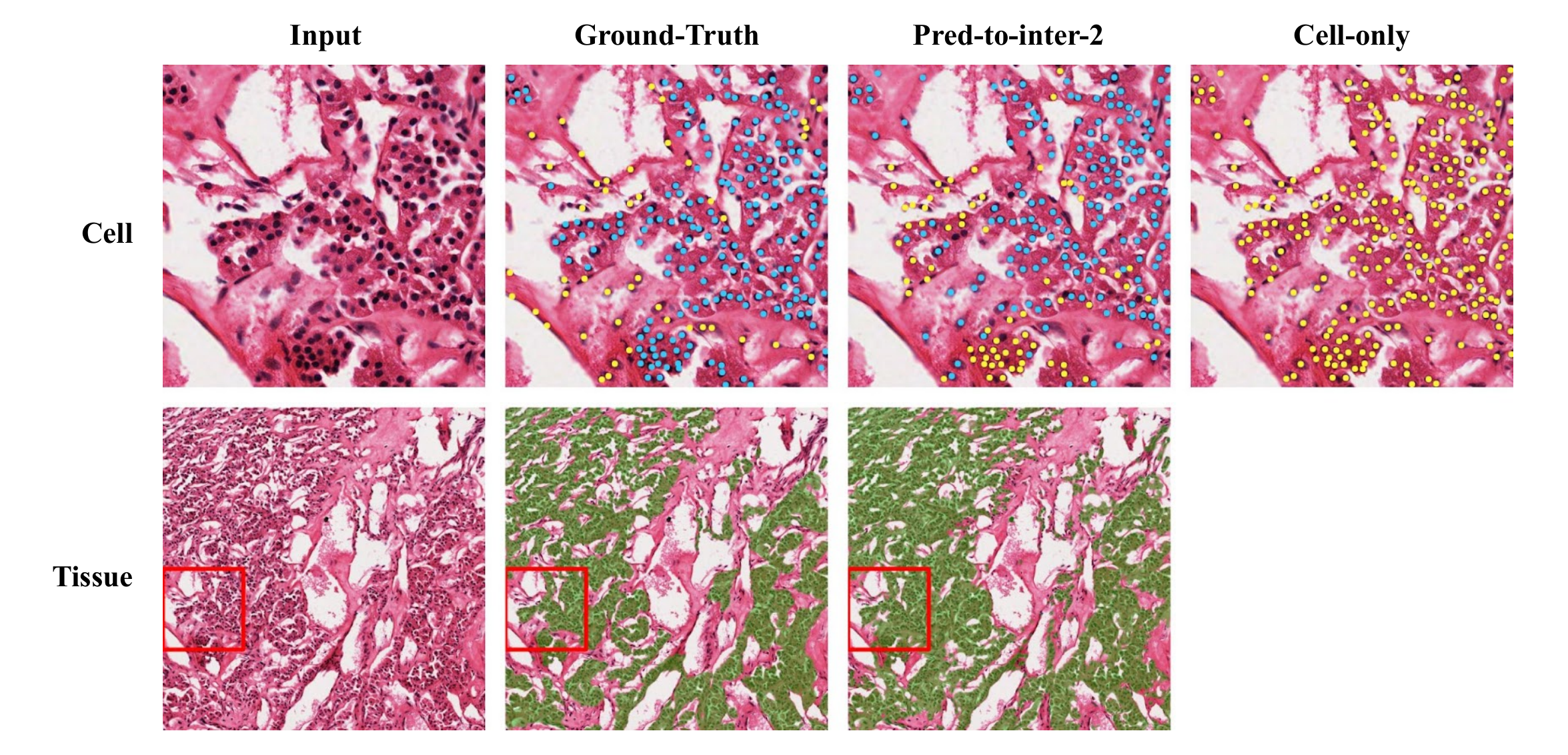}
    \vspace{-3mm}
    \caption{\textbf{Qualitative comparison between the \textit{Pred-to-inter-2} and \textit{Cell-only} models}. \textit{Pred-to-inter-2} shows enhanced detection results aided by understanding the tissue environment from a broader context (blue: tumor cells, yellow: background cells, green: cancer area).}
\label{fig:qualitative}
\vspace{-4mm}
\end{figure*}

\subsection{Dataset Ablation Study}
Two data components in \ours contribute to better leveraging cell-tissue relationships for the task of cell detection: (1) the large FoV patches $x_l$, and (2) the corresponding tissue annotations $y_l^t$ that allows us to have a tissue segmentation objective and a multi-task learning setting.
To verify the effectiveness of each data component, we conduct an ablation study where each component is evaluated separately while keeping the model architecture fixed.
First, if only the tissue segmentation label is provided without a large FoV patch, the tissue branch takes $x_s$ as an input and it is trained with both the tissue and the cell detection objectives.
Second, if the large FoV patch is given without the corresponding tissue label, the tissue branch takes $x_l$ as an input and the model is only trained with the cell detection objective.

The results of the ablation study are shown in \autoref{tab:ablation}. We observe that each of the individual components on their own lead to improved performance. Hence, we can conclude that a large FoV patch and tissue segmentation label can contribute to better cell detection by providing a broader context and encoding the cell-tissue relationship, respectively.
Moreover, when utilized simultaneously, we observe that both components work synergetically, leading to an even better performance improvement.
We also found that the tissue branch generates better tissue predictions when trained with a large FoV. Thus, providing accurate tissue information to the cell branch is important for boosting performance. Please, refer to the \supple for a detailed tissue performance comparison.
Finally, this study shows evidence that the patch configuration defined in \ours is suitable for cell-tissue co-training, with a smaller FoV for cell detection $(x_s, y_s^c)$ and a larger FoV for tissue segmentation $(x_l, y_l^t)$.

\subsection{Qualitative Analysis} \label{subsec:qualitative}
In this section, we visualize the cell predictions of the \textit{Cell-only} and \textit{Pred-to-inter-2} models.
We select the \textit{Pred-to-inter-2} model, as it is the overall best-performing model in our experiments.
In general, when compared to other background cells, tumor cells have the following characteristics: large size and irregular shape. However, cancer is heterogeneous and this is not always the case.
Indeed, in \autoref{fig:qualitative}, most of the cells are small and have a regular round shape. Based on these appearances, and without a larger context, those cells can be easily misclassified as background cells, which is the case of the \textit{Cell-only} model.
On the other hand, \textit{Pred-to-inter-2} shows a more accurate prediction by correctly understanding the cancer area in large FoV regions.
This implies that \textit{Pred-to-inter-2} indeed considers both the morphology of cells and the tissue context, while \textit{Cell-only} relies on the cells' morphology alone. 

\section{Conclusion}
\label{sec:conclusion}
In histopathology, there is a hierarchical organization and interrelationship between cells and tissues. Hence, we hypothesize that it can be leveraged to improve the performance of cell detection tasks. Throughout this paper, we have shown evidence that such cell-tissue relationships exist across multiple setups. Indeed, we observe improvements when utilizing large FoVs with corresponding tissue annotations together with cell annotations in a multi-task learning approach, using simple neural architecture designs. Although the cell detection task clearly benefits from the multi-task learning framework, the improvement on the tissue segmentation task was not yet investigated, and we leave this topic as a future direction. We hope this work, especially the \ours dataset, initiates new and promising research directions in CPATH to further study the cell-tissue relationships and improve cell detection and tissue segmentation.

{\small
\bibliographystyle{ieee_fullname}
\bibliography{egbib}
}

\clearpage

\section*{Appendix}
\appendix
\counterwithin{figure}{section}
\counterwithin{table}{section}

\noindent\textbf{Note:} We use \textcolor{blue}{blue} color to refer to section numbers \textbf{in the main paper}. All \textcolor{red}{red} and \textcolor{green}{green} characters refer to figures, tables, and citations in this supplementary material.
\vspace{0.3cm}

\noindent\textbf{Overview.} This supplementary material includes further information regarding the implementation details, results, and datasets discussed in the main paper, and summarized as,

\begin{itemize}
    \item We detail how the cell detection task is posed as a segmentation task, and how cell detection and tissue segmentation tasks are handled simultaneously. 

    \item We show tissue segmentation results to provide more insights on how large FoVs and the corresponding tissue label improve cell detection performance, as discussed in \textcolor{blue}{Subsection 6.4} and \textcolor{blue}{Tab. 7}.
    \item We share qualitative results comparing the \textit{Cell-only} and \textit{Pred-to-inter-2} models in OCELOT and CARP.

    \item Finally, regarding the datasets, we provide the amount of annotated cells and tissue pixels per data subset of OCELOT in \autoref{tab:label_stats_ocelot2}. For TIGER \cite{tiger}, we describe how the dataset is pre-processed in order to be used in our experiments.
\end{itemize}

\section{Annotation Protocol \textcolor{blue}{(Section 3.1)}}

\paragraph{Annotation rules.} For cell patches, annotators were asked to annotate the center point of each cell. For tissue patches, annotators drew contours as accurately as possible.

\paragraph{Consensus strategy.} All data were annotated by board-certified pathologists. Each tissue patch is annotated by a single pathologist. Each cell patch is annotated by three pathologists with the following consensus strategy. First, two pathologists annotate the same cell patch independently. Then, the third pathologist merges the two annotations taking the discrepancies into account. This strategy was specifically designed to reduce the naturally high inter-rater variability when annotating cells.
\begin{table}[t]
\small
\addtolength{\leftskip} {-2cm}
\addtolength{\rightskip}{-4cm}
\setlength{\tabcolsep}{0.2em}
\centering
{
\begin{tabular}{lrrrclrrr}
\cmidrule[\heavyrulewidth]{1-4}\cmidrule[\heavyrulewidth]{6-9}
\multirow{2}[2]{*}{\textbf{}}  & \multicolumn{3}{c}{\textbf{\# Pixels}}                                        & \enspace & \multirow{2}[2]{*}{\textbf{}} & \multicolumn{3}{c}{\textbf{\# Cell}} \\ \cmidrule(l){2-4}\cmidrule(l){7-9}

& \multicolumn{1}{c}{\textbf{Train}}   & \multicolumn{1}{c}{\textbf{Val}} & \multicolumn{1}{c}{\textbf{Test}} &   &
   & \multicolumn{1}{c}{\textbf{Train}} & \multicolumn{1}{c}{\textbf{Val}} & \multicolumn{1}{c}{\textbf{Test}}\\ \cmidrule{1-4}\cmidrule{6-9}
\textit{BG}\enspace  & 237.4M  & 79.1M  & 71.6M  & &
\textit{TC} & 43.8K  & 16.3K & 12.9K \\
\textit{CA}  & 171.0M  & 57.8M & 58.8M &  & 
\textit{BC}  & 23.6K & 8.4K  & 9.7K  \\

\textit{UNK} & 17.3M & 6.7M & 6.1M &  & 
{} & {}  & {} & {} \\  
\cmidrule[\heavyrulewidth]{1-4}\cmidrule[\heavyrulewidth]{6-9}
\textbf{Total} & 425.7M & 143.6M & 136.3M &  & 
\textbf{Total} & 67.4K  & 24.7K & 22.6K \\  
\cmidrule[\heavyrulewidth]{1-4}\cmidrule[\heavyrulewidth]{6-9}
\multicolumn{4}{c}{(a) Tissue Annotations} & & \multicolumn{4}{c}{(b) Cell Annotations} \\
\end{tabular}
}
\caption{\textbf{Annotation statistics of OCELOT.} In (a), BG, CA, and UNK denote Background, Cancer Area, and Unknown tissue classes, respectively. The pixel counts were computed from the down-sampled tissue patches (1024 $\times$ 1024). In (b), TC and BC denote Tumor cells and Background cells, respectively. 
}
\label{tab:label_stats_ocelot2}
\end{table}

\section{Implementation Details \textcolor{blue}{(Section 5.1)}}

\paragraph{Cell detection as segmentation.}
We define the cell detection task as a segmentation one, similarly to \cite{SwiderskaChadaj2019LearningTD}. At training time, we provide the cell labels as a segmentation map by drawing a disk centered on each cell point annotation. We use a fixed radius of 1.4 $\mu\text{m}$, corresponding to 7 pixels at a resolution of 0.2 Microns-per-Pixel (MPP). Then, we assign the value of each pixel within each disk to the corresponding cell label, e.g., 1 for TC and 2 for BC in OCELOT; 0 for the remaining background pixels. We utilize the Dice loss \cite{dice} for both cell and tissue branches, which is a widely used loss function for semantic segmentation.

At inference time, we post-process the probabilistic cell segmentation map, i.e., the output of the cell branch, to obtain a set of points, corresponding to the detection of the cells.
To that end, we apply \verb|skimage.feature.peak_local_max|\footnote{\scriptsize\url{https://scikit-image.org/docs/stable/api/skimage.feature.html\#skimage.feature.peak\_local\_max}} on the cell segmentation map to get the set of predicted points (cells). 
Lastly, we retrieve the class probability values of each cell from the segmentation maps and determine their class through \verb|argmax|. The class probability is used as the confidence score. 

\paragraph{Data augmentation.} 
During training, five data augmentations are randomly applied, including three photometric (gaussian blur, gaussian noise, color jitter) and two geometric (horizontal flipping, rotation by a multiple of 90$^{\circ}$) transformations.
In the case of geometric transformations, we apply the same transformation for cell and tissue patches within a pair to maintain the physical alignment between them (e.g. 90$^{\circ}$ for both cell and tissue patches).

\paragraph{Learning rate and dropout for cell and tissue branches.}
During experiments, we find that the convergence speeds of the cell detection and tissue segmentation tasks are different. The cell branch starts overfitting while the tissue branch is still learning. 
To address this behavior, we use different dropout probabilities and learning rates (LRs) for each branch.
In the case of dropout, a fixed probability value of 0.1 is used for the tissue branch. 
Conversely, we tune the cell branch by performing a grid search with 3 dropout probability values: 0.1, 0.3, and 0.5.
Note that the dropout layer is added at the end of each ResNet block. We use spatial dropout \cite{tompson2015efficient}.
In the case of the LR, while searching for the best hyper-parameter values, we constrain the LR of the cell branch to be the same or half of the LR of the tissue branch. This constraint is applied to reduce the search space. 

\section{More Cell Detection Baselines \textcolor{blue}{(Section 6.1)}}
We provide more cell detection baselines (U-Net \cite{unet} and MFoVCENet \cite{10.1007/978-3-030-59722-1_37}) on the OCELOT dataset. MFoVCE-Net is a strong baseline that further utilizes a large FoV patch as an input, but not a corresponding tissue annotation. \autoref{tab:more-baseline} shows that the proposed \textit{Pred-to-inter-2} model still outperforms all the baselines by a large margin. This emphasizes the importance of the additional larger FoV input and associated tissue label. In addition, the U-Net architecture shows lower performance than DeepLabV3+ \cite{deeplabv3plus2018}.

\setcounter{section}{3}
\setcounter{table}{0}
\begin{table}[t]
\small
\centering
\setlength{\tabcolsep}{0.5em}
\begin{tabular}{llrcl}
\toprule
\textbf{Method} & \textbf{Architecture} & \multicolumn{1}{c}{\textbf{Val}} & \multicolumn{1}{c}{\textbf{Test}}\\ 
\midrule
Cell-only & DeepLabV3+ \cite{deeplabv3plus2018} & \score{68.87}{1.76} & \score{64.44}{1.82} \\
Cell-only & U-Net & \score{67.75}{1.42} & \score{63.46}{4.59} \\
Cell-only$\dagger$ & MFoVCE-Net \cite{10.1007/978-3-030-59722-1_37} & \score{69.14}{0.52} & \score{67.12}{1.96}\\
\midrule
\textit{Pred-to-inter-2} & DeepLabV3+ \cite{deeplabv3plus2018} & \score{72.68}{1.58} & \score{71.23}{0.96}\\
\bottomrule
\end{tabular}
\vskip -2mm
\caption{\textbf{More cell detection baselines.} Comparison with various cell detection methods. $\dagger$ denotes that a large FoV patch is also utilized as an input.
}
\label{tab:more-baseline}
\vspace*{-1mm}
\end{table}

\setcounter{section}{4}
\setcounter{table}{0}
\begin{table*}[th!]
{
\addtolength{\leftskip} {-2cm}
\addtolength{\rightskip}{-2cm}
\setlength{\tabcolsep}{0.3em}
\centering
\resizebox{0.6\linewidth}{!}
{
    \begin{tabular}{ccrrrr}
    \toprule
    \multicolumn{2}{c}{Further utilize} & \multicolumn{2}{c}{\textit{Cell (mF1)}} & \multicolumn{2}{c}{\textit{Tissue (mIoU)}} \\ \cmidrule(lr){3-4}\cmidrule(lr){5-6}
    Tissue label    & Large FoV     & \multicolumn{1}{c}{Val}   & \multicolumn{1}{c}{Test}  & \multicolumn{1}{c}{Val}   & \multicolumn{1}{c}{Test}  \\ \midrule
    -               & -             & \score{68.87}{1.76}       & \score{64.45}{1.82}       & N/A       & N/A  \\
    \checkmark      & -             & \score{70.36}{1.34}       & \score{66.58}{2.02}       & \score{75.27}{3.10}       & \score{73.75}{3.74} \\
    -               & \checkmark    & \score{69.61}{1.59}       & \score{66.22}{1.53}                & N/A       & N/A \\
    \checkmark      & \checkmark    &\textbf{\score{72.30}{0.73}} & \textbf{\score{68.91}{2.52}} & \textbf{\score{77.48}{1.96}} & \textbf{\score{81.97}{1.75}} \\
    \bottomrule
  \end{tabular}
}
\vskip -1mm
\caption{\textbf{Ablation study on the tissue segmentation performance.} 
Tissue segmentation performance is further reported beyond the ablation study in \textcolor{blue}{Tab. 7}.
The first row corresponds to the \textit{Cell-only} model, and the third row includes the tissue branch with large FoV, but without tissue supervision.
Since the models in two of the rows do not consider the tissue labels, we denote their performance as N/A.
Note that the \textit{Feature-sharing} model and the OCELOT dataset are used.}
\label{tab:ablation_supple}
\vspace{-3mm}
}
\end{table*}

\setcounter{section}{3}
\section{Ablation Study: Tissue Performance \textcolor{blue}{(Section 6.4)}}

Through the ablation study in the \textcolor{blue}{Tab. 7} of the main document, we observe improvements in cell detection performance by utilizing a large FoV or tissue segmentation label. Moreover, utilizing both components simultaneously shows synergy, leading to an even better performance improvement. 
In \autoref{tab:ablation_supple}, we investigate the tissue segmentation performance to better understand the reason for such synergy. 
By comparing the second and last rows in \autoref{tab:ablation_supple}, we observe that training with large input/label FoV tissue results in a better tissue model, which achieves higher mIoU in both validation and test sets. Therefore, the cell detection performance boost can be justified by the fact that the tissue model shares more accurate tissue information to the cell branch.

\section{Qualitative Results \textcolor{blue}{(Section 6.5)}}
\begin{figure*}[t]
    \centering
    \vspace*{-2mm}
    \includegraphics[width=0.85\linewidth]{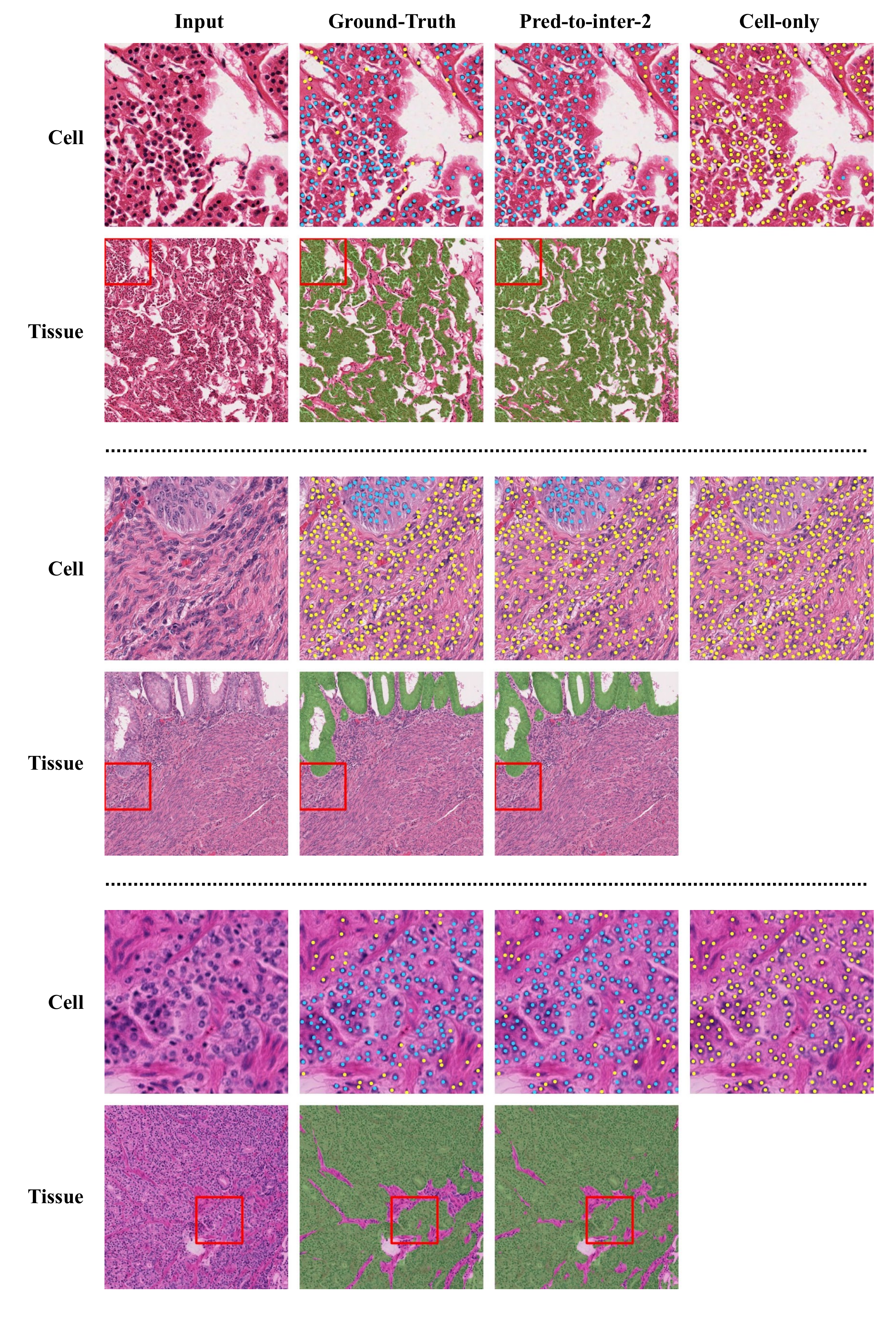}
    \vspace{-3mm}
    \caption{\textbf{Qualitative results - OCELOT}. 
        The \emph{Pred-to-inter-2} model can correct the mistakes of the \emph{Cell-only} model by incorporating tissue prediction information during cell prediction.
        The colors represent the following classes:
        \textcolor{cyan}{$\mdlgblkcircle$} Tumor Cells (TC), 
        \textcolor{yellow}{$\mdlgblkcircle$} Background Cells (BC), and
        \textcolor{ForestGreen}{$\mdlgblksquare$} Cancer Area (CA).
    }
\label{fig:supple_ours}
\vspace{-4mm}
\end{figure*}

\begin{figure*}[t]
    \centering
    \vspace*{-2mm}
    \includegraphics[width=0.85\linewidth]{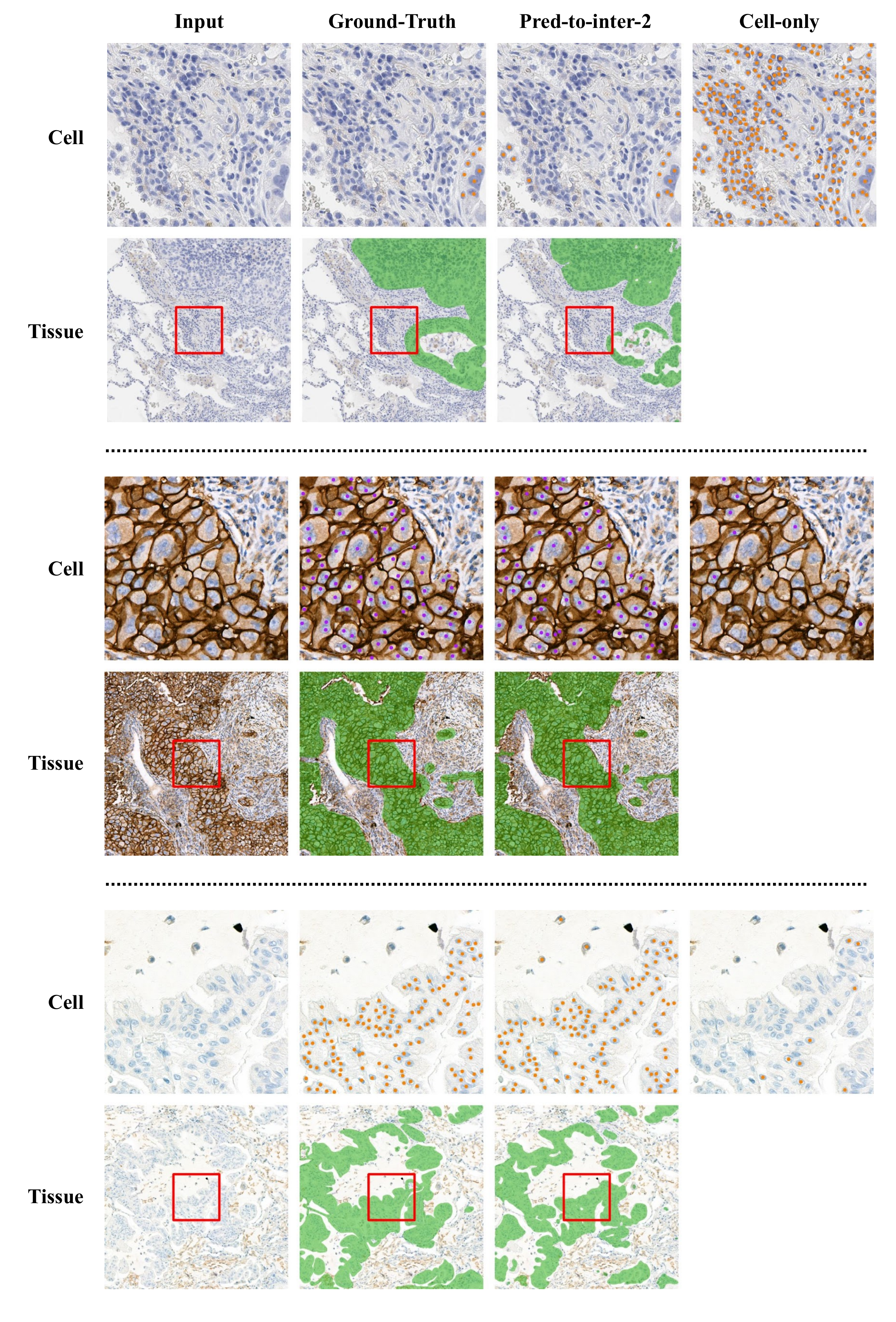}
    \vspace{-3mm}
    \caption{\textbf{Qualitative results - CARP}.
        The \emph{Pred-to-inter-2} model can correct the mistakes of the \emph{Cell-only} model by incorporating tissue prediction information during cell prediction.
        The colors represent the following classes:
        \textcolor{Orchid}{$\mdlgblkcircle$} PD-L1 positive tumor cells (TC+),
        \textcolor{orange}{$\mdlgblkcircle$} PD-L1 negative tumor cells (TC-), and
        \textcolor{ForestGreen}{$\mdlgblksquare$} Cancer Area (CA).
    }
\label{fig:supple_carp}
\vspace{-4mm}
\end{figure*}

We provide more examples for qualitative comparison between \textit{Cell-only} and \textit{Pred-to-inter-2} models. Visualizations of OCELOT can be found in \autoref{fig:supple_ours} and CARP in \autoref{fig:supple_carp}. We use a different color scheme for each figure since each dataset is based on different staining methods. The color scheme can be found in the captions.

\section{Details about TIGER \textcolor{blue}{(Section 6.1)}}
\paragraph{Annotations.} There is a single class annotation for the cell task, namely, lymphocyte cells. In contrast, 7 classes are considered for the tissue task: \textit{Invasive Tumor}, \textit{Tumor-associated Stroma}, \textit{In-situ Tumor}, \textit{Healthy Glands}, \textit{Necrosis not in-situ}, \textit{Inflamed Stroma}, and \textit{Rest}. In addition, TIGER considers the tissue class \textit{Excluded}, which has the same role as \textit{Unknown} in OCELOT. 

Based on the statistics in \autoref{tab:tiger_class_ratio}, we observe that most of the lymphocyte cells are located within stroma tissue areas, i.e., \textit{Tumor-associated Stroma} and \textit{Inflamed Stroma.} 
Also, the tissue annotations suffer from severe class imbalance. In fact, the frequencies of 4 out of 7 classes are lower than 5\%. 
To make the tissue task more straightforward, while maintaining the interrelation between lymphocyte cells and stroma tissue, we remap the tissue classes as follows: \textit{Tumor-associated Stroma} and \textit{Inflamed Stroma} are grouped into the \textit{Stroma} (\textit{ST}) class, and the remaining labels are remapped to \textit{BG} class.
Note that the main goal of this work is to explore cell-tissue relationships for improving the cell detection task, not to tackle the tissue segmentation task explicitly.

\begin{table}[t]
\small
\centering
\setlength{\tabcolsep}{0.2em}

\begin{tabular}{lrrcl}
\toprule
 \multicolumn{1}{c}{\textbf{Tissue Class Name}} & \multicolumn{1}{c}{\textbf{Class Ratio}} &
  \multicolumn{1}{c}{\quad\textbf{LC in Tissue Ratio}}
 \\ 
\midrule

Invasive Tumor & \quad27.11 \% & \quad4.14 \% \\
Tumor-associated Stroma  & \quad27.48 \% & \quad30.36 \% \\
In-situ Tumor & \quad4.86 \% & \quad0.14 \% \\
Healthy Glands  & \quad3.05 \% & \quad0.66 \% \\
Necrosis not in-situ & \quad1.48 \% & \quad0.09 \% \\
Inflamed Stroma & \quad3.31 \% & \quad58.98 \% \\
Rest & \quad31.67 \% & \quad4.65 \% \\
Excluded & \quad1.04 \% & \quad0.98 \% \\

\midrule

\end{tabular}
\vskip -2mm
\caption{\textbf{TIGER class ratio}. LC denotes lymphocyte cell. We observe that most LCs are located within stroma regions.}
\label{tab:tiger_class_ratio}
\end{table}
\begin{table}[t]
\small
\centering
\setlength{\tabcolsep}{0.2em}

\begin{tabular}{lrrrrcl}
\toprule
 \multicolumn{1}{c}{\textbf{Dataset}} & \multicolumn{1}{c}{\textbf{Cell FoV}} &
  \multicolumn{1}{c}{\textbf{Tissue FoV}} &
  \multicolumn{1}{c}{\quad\textbf{MPP}} &
  \multicolumn{1}{c}{\quad\textbf{\# of patch pairs}} 
 \\ 
\midrule

OCELOT & \quad1024 & \quad4096 & \quad$\sim$0.2 & 673 \\
CARP & \quad1024 & \quad4096 & \quad$\sim$0.2 & 6,480 \\
TIGER$^*$ & \quad128 & \quad512 & \quad$\sim$0.5 & 9,888 \\

\midrule

\end{tabular}
\vskip -2mm
\caption{\textbf{Dataset comparison after TIGER pre-processing}. TIGER$^*$ denotes the TIGER dataset after pre-processing. \textit{\# of patch pairs} includes all the samples in \textit{training}, \textit{validation}, and \textit{test}.}
\label{tab:supple_dataset_comparison}
\vspace{-3mm}
\end{table}
\paragraph{Data pre-processing.} A pre-processing step is necessary for the TIGER dataset due to the inconsistent annotated ROI sizes for both cell and tissue samples. We can identify two different subsets in TIGER: 1) the sample pairs from the TCGA \cite{HUTTER2018283} database, and 2) the pairs from other sources, which we denote as non-TCGA pairs. On one hand, TCGA samples are composed of large annotated tissue patches that contain several smaller cell annotated ROIs within their region.
The number of cell ROIs per sample highly varies, reaching up to 58.
These cell ROIs are variable in size and most of them are smaller than 256 $\times$ 256 pixels. On the other hand, non-TCGA samples have a complete overlap between the cell and tissue patches, and the size of these patches is larger than 512 $\times$ 512 pixels. 

To maximize the amount of usable cell-tissue sample pairs, while maintaining the 4 times FoV difference across the cell and tissue tasks (as done in OCELOT and CARP), we define the cell and tissue FoVs to be 128 $\times$ 128 and 512 $\times$ 512 pixels, respectively. Note that the image patch size is considerably smaller than in OCELOT and CARP mainly because of the limited size of cell ROIs in TCGA samples. In addition, the pre-processing step is implemented differently according to the data source; TCGA samples (see \autoref{alg:tcga_preproc}) and non-TCGA samples (see \autoref{alg:nontcga_preproc}). As a result of this pre-processing step, each non-TCGA tissue patch is paired to $4^2$ different cell sub-patches. In contrast, for each cell ROI in TCGA, there can be up to $4^2$ surrounding tissue patches. Please, refer to \autoref{tab:supple_dataset_comparison} for a comparison of the statistics across OCELOT, CARP, and the pre-processed TIGER datasets.
\vspace{-1mm}

\begin{algorithm*}
    \caption{TIGER pre-processing step for TCGA samples } 
    \label{alg:tcga_preproc}
    \begin{algorithmic}[1]
        \State \textbf{Input} $\mathcal{D}_{TCGA}$ \Comment{TCGA dataset}
        \State \textbf{Output} $\mathcal{D}_{proc}$ \Comment{Pre-processed dataset}
        \State $Sz_c$, $Sz_t$ $\gets$ 128, 512  \Comment{Cell and tissue patch sizes, respectively}
        \State $\mathcal{D}_{proc} \gets [\hspace{0.25em}]$ \Comment{Initialize pre-processed dataset to empty list}
        \ForAll{$(img_t, imgs_c^{ROIs}, masks_t$, $bboxes_c^{ROIs}$, $info^{ROIs})$  \textbf{in} $\mathcal{D}_{TCGA}$} \Comment{Loop over the dataset}
        
            \State $H_t, W_t$ $\gets$ Size($img_t$) \Comment{Tissue image dimension}
            \ForAll{$(img_c$ ,$bboxes_c$, $info_c)$ \textbf{in} $(imgs_c^{ROIs},  bboxes_c^{ROIs}, info^{ROIs})$} \Comment{Loop over cell ROIs in a sample}
                \State $H_c, W_c$ $\gets$ Size($img_c$) \Comment{Cell image dimension}
                \State $y_c$, $x_c$ $\gets$ GetROILocation($info_c$) \Comment{Getting the top-left coordinates of the cell ROI}
                \If{$W_c < Sz_c$ or $H_c < Sz_c$} \Comment{Ignore small cell ROIs}
                    \State \textbf{continue}  
                \EndIf
                \State $img_c \gets$ Crop($img_c$, ($0,0$), $Sz_c$) \Comment{Cropping cell ROI from the top-left corner (0,0) and size $Sz_c$} 
                \State $bboxes_c$ $\gets$ FilterBboxes($bboxes_c$, $y_c$, $x_c$, $Sz_c$) \Comment{Removing cell bounding boxes due to previous cropping}
                \ForAll{$(i_t,j_t) \in [0..Sz_t/Sz_c] \times [0..Sz_t/Sz_c]$} \Comment{Loop over 16 surrounding tissue patches per cell ROI}
                    \State $y_t, x_t \gets y_c - i_t \cdot Sz_c, x_c - j_t \cdot Sz_c$ \Comment{Defining surrounding tissue coordinates}
                    \If{CheckTissueExcelsImg($y_t, x_t, Sz_t, H_t, W_t$)} \Comment{Ignore tissue patches exceeding the image}
                        \State \textbf{continue} 
                    \EndIf
                    \State $curImg_t \gets$ Crop($img_t$, ($y_t,x_t$), $Sz_t$) \Comment{Cropping tissue surrounding patch}
                    \State $curMask_t \gets$ Crop($mask_t$, ($y_t,x_t$), $Sz_t$)
                    \State $resImg_t$ $\gets$ Resize($curImg_t$, ($Sz_c$,$ Sz_c$)) \Comment{Matching tissue to cell size}
                    \State $resMask_t$ $\gets$ Resize($curMask_t$, ($Sz_c$, $Sz_c$))
                    \State AppendTo($\mathcal{D}_{proc}$, ($img_c$, $bboxes_c$, $resImg_t$, $resMask_t$)) \Comment{Save sample in pre-processed dataset}
                \EndFor
            \EndFor
        \EndFor
    \end{algorithmic}
\end{algorithm*}

\begin{algorithm*}
    \caption{TIGER pre-processing step for non-TCGA samples}
    \label{alg:nontcga_preproc}
    \begin{algorithmic}[1]
        \State \textbf{Input} $\mathcal{D}_{nonTCGA}$ \Comment{Non-TCGA dataset}
        \State \textbf{Output} $\mathcal{D}_{proc}$ \Comment{Pre-processed dataset}
        \State $\mathcal{D}_{proc} \gets [\hspace{0.25em}] $
        \State $Sz_c$, $Sz_t$ $\gets$ 128, 512  \Comment{Cell and tissue patch sizes, respectively}
        \ForAll{$(img, mask, bboxes)$ \textbf{in} $\mathcal{D}_{nonTCGA}$} \Comment{Looping over perfectly overlapping cell-tissue images}
            \State $H$, $W$ $\gets$ Size($img$) \Comment{Image dimension}
            \If{$H > 1024$ and $W > 1024$} \Comment{Consider only large samples}
                \ForAll{$(i_t,j_t) \in  [0..H_t/Sz_t]\times[0..W_t/Sz_t]$} \Comment{Tissue 2D patch loop}
                        \State $y_t, x_t \gets i_t \cdot Sz_t,  j_t \cdot Sz_t$ \Comment{Define top-left coordinates of the tissue patch}
                        \State $img_t \gets$ Crop($img$, ($y_t,x_t$), $Sz_t$) \Comment{Cropping tissue patch}
                        \State $mask_t \gets$ Crop($mask$, ($y_t,x_t$), $Sz_t$)
                        \State $resImg_t \gets$ Resize($img_t$, ($Sz_c$, $Sz_c$))  \Comment{Matching tissue to cell size}
                        \State $resMask_t \gets$ Resize($mask_t$, ($Sz_c$, $Sz_c$))
                        \ForAll{$(i_c,j_c) \in [0..Sz_t/Sz_c] \times [0..Sz_t/Sz_c]$} \Comment{Cell 2D sub-patch loop}
                                \State $y_c, x_c \gets i_c \cdot Sz_c, j_c \cdot Sz_c$ \Comment{Define cell patch coordinates}
                                \State $img_c \gets$ Crop($img_t$, ($y_c,x_c$), $Sz_t$) \Comment{Cropping cell sub-patch in tissue patch}
                                \State $bboxInPatch \gets$ FilterBboxes($bboxes, y_c + y_t, x_c + y_t, Sz_c$) \Comment{Removing bboxes off the cell sub-patch}
                                \State AppendTo($\mathcal{D}_{proc}$, ($img_c$, $bboxes_c$, $resImg_t$, $resMask_t$)) \Comment{Save sample in pre-processed dataset}
                        \EndFor
                \EndFor
            \EndIf
        \EndFor
    \end{algorithmic}
\end{algorithm*}

\end{document}